\documentclass[12pt]{article}
\newcommand{\bmx}{\mbox{\boldmath $x$}}

\newcommand{\bbeta}{\mbox{\boldmath $\beta$}}
\newcommand{\bgamma}{\mbox{\boldmath $\gamma$}}

\newcommand{\bmz}{\mbox{\boldmath $Z$}}

\usepackage[dvips]{graphics}
\usepackage{amsmath}
\usepackage[round]{natbib}
\usepackage{url}
\usepackage{hyperref}
\usepackage{hyperref}
\usepackage{breakurl}
\usepackage{footmisc}
\usepackage{xspace}
\usepackage[nottoc]{tocbibind}
\usepackage{booktabs, array}
\usepackage{longtable,rotating}
\usepackage{nomencl}
\usepackage{subfigure}
\usepackage{fmtcount}
\usepackage{fancyhdr}
\usepackage{bbding}
\usepackage{pifont}
\usepackage{wasysym}
\usepackage{amssymb}

\begin{document}

\title{\bf  Statistical methods for body mass index: a selective review of the literature}
\author{Keming Yu
\thanks{{\it Address for correspondence:}  Department of Mathematics, Brunel University London, UK. \newline E-mail: keming.yu@brunel.ac.uk},    Rahim Alhamzawi,  Frauke Becker and Joanne Lord\\
{\em\small   }
}
\date{}
\maketitle

\noindent {\bf Summary.} Obesity rates have been increasing over recent decades, causing significant concern among policy makers. Excess body fat, commonly measured by body mass index (BMI), is a major risk factor for several common disorders including diabetes and cardiovascular disease, placing a substantial burden on health care systems. 
To guide effective public health action, we need to understand the complex system of intercorrelated influences on BMI. This paper will review both classical and modern statistical methods for BMI analysis, highlighting that most of the classical methods are simple and easy to implement but ignore the complexity of data and structure, whereas modern methods do take complexity into consideration but can be difficult to implement. A series of case studies are presented to illustrate these methods and some potentially useful new models are suggested.


\section{Introduction}

Obesity is considered one of the most important medical and public health problems of our time (\cite{ells2009preventing}). Excess body fat has been identified as a major risk factor for several common disorders including diabetes and cardiovascular diseases and  imposes a substantial burden on health care systems. Recently, the American Medical Association classified obesity itself as a disease
(\url{http://www.forbes.com/sites/brucejapsen/2013/06/18/ama-backs-disease}\\
\url{-classification-for-obesity/}), increasing the focus on its importance as a public health concern.

Obesity can be measured in various ways. Body mass index (BMI) is the most commonly used measure of relative weight
. It can be used both at individual level to asses body weight in a clinical setting and at population level where it would be impractical or too expensive to measure (excess) body fat accurately and consistently. Based on individual height and weight, BMI is defined as body weight measured in kilograms 
divided by the square of height in meters
:
\[ BMI=\frac{mass(kg)}{height(m)^2} 
. \]
Applications of BMI often rely on the classification of `healthy' and `unhealthy' segments of the BMI distribution (W. H. \cite{world2000obesity}).
The  contribution of statistical methods to understanding BMI is in collection, organization, analysis and presentation of data as well as the interpretation of results and prediction. The use of statistical methods in 
BMI-related research may provide support to:
\begin{enumerate}
  \item
	Investigate factors associated with BMI, and identify the relationship between BMI and correlated (potentially causal) factors.

Although BMI itself is easy to calculate, the system of underlying contributing factors and their intercorrelation is multifaceted. At the individual level, obesity is caused by a continuously positive energy balance, when more calories are consumed than expended. However, the influences driving individual choices which affect the energy balance are highly complex. Within the UK Government's Foresight Programme, a system map was developed that describes the obesogenic environment of interacting influences on weight gain, without identifying any single dominating factor (Vandenbroeck, Goossens and Clemens, 2007). 
In addition to food and physical activity choices, these influences include biological and medical traits, social and psychological components, as well as effects from the built environment and infrastructure.


\item	  Investigate how BMI may contribute to morbidity and mortality from various related diseases. In a public health context, BMI can be used as a predictor of unhealthy body weight and related disease at a population level. Applying statistical methods to analyse BMI data may help to identify the correlations between obesity, health characteristics and influencing factors. For example, obesity and inactivity are known risk factors for type 2 diabetes 
and evidence has been published on the associations between BMI and (i) different types of cancer (\citealp{renehan2008body}; \citealp{larsson2008excess}) and (ii) risk of gestational diabetes mellitus (GDM) 
\citep{torloni2009prepregnancy}.

\item Explore the classification of BMI and address uncertainty.

BMI is a sufficiently good proxy to capture obesity on population level, but does not provide a direct indication of the distribution of body fat. Statistical methods can provide a way to investigate the distribution of body fat based on BMI datasets. According to  \cite{dinsdale2011simple}, the British 1990 growth reference (UK90) is recommended for population monitoring and clinical assessment in children aged four years and over providing centile curves for BMI 
as a norm against which changes in individual measurements can be monitored. 

\item Inform  policy making process by  evaluation and prediction.

A better understanding of how individual characteristics, choices and influences affect body weight and how excessive body fat is associated with increased risk of disease and mortality is essential for the identification of cost-effective interventions. Various statistical designs and methods can be used for the analysis of 
BMI improving the understanding of (i) how to target influencing factors in order to change BMI, and (ii) how to predict variation in BMI based on specific factors.


\end{enumerate}

\subsection{Overview of statistical methods}

Statistical methods  to identify patterns and trends within large datasets  are now integral to the development of scientific research in biological and process modelling, personalized healthcare, pharmacology, health economics, and public health policy. In addition to general concerns such as environmental influences, genetics and disease prevalence, statistical issues that necessitate more advanced models for the analysis of BMI data include skewness, outliers and non-response values.

Table 1 provides a summary of a range of statistical methods and models 
that can be applied to address these problems.
\begin{table}[ht]
\caption{{\bf Statistical Methods or Models}} 
\centering 
\begin{tabular}{l|| l }
\hline\hline
{\bf Methods or models}&{\bf Sections} \\
\hline\hline
 Correlation Analysis or Effect Analysis & 2, 4 \\ \hline
 Multiple linear regression &2, 3, 4\\ \hline
 Random or mixed effect regression& 2, 3\\ \hline
Quantile regression & 4 \\ \hline
Ignorable likelihood  and multiple imputation methods & 5\\ \hline
Longitudinal models  and Methods &2, 6\\ \hline
Bayesian methods &4, 5\\ \hline
Moment methods and estimation equations &6\\ \hline
Dimension reduction methods  &2, 6\\ \hline
  \end{tabular}
\label{table:nonlin} 
\end{table}
Most of these statistical models could be summarized by a single equation 
:
\begin{equation}
h(m(Y))=f_1(\bbeta'\, \bmx)\,+\,f_2(\bgamma'\, \bmz)\,+\,f_3(t)+\epsilon,
\end{equation}
where $Y$ is the response variable such as BMI or disease incidence, $m(Y)$ is a characteristic of $Y$ such as the mean $E(Y)$, variance $Var(Y)$ or quantile $Q(Y)$, and $h(.)$ defines a link function such as a logarithm function or logistic function.
$\bbeta' \bmx$ specifies a linear combination of explanatory variables  $\bmx$, $\bgamma' \bmz$ is a collection of random factors taking grouping of data into account. $\epsilon$ is the model error or random noise.
Each of $f_1(.)$, $f_2(.)$ and $f_3(.)$  could be a linear function where $f(\bbeta'\bmx)=\bbeta'\bmx$, a nonlinear function such as  $f_2(\bgamma'\bmz)=I[\bgamma' \bmz>30]$ , a nonparametric regression function such as an unknown smooth function $f_3(t)$ over time $t$, or even a time series such as an autoregressive model.


{\bf \subsection{Review of the literature}}

We conducted a literature review to identify statistical methods that could be applied to the above problems. We focussed on statistical methods and models which have been used in the analysis of BMI, extending classical methods and applying more sophisticated models. The search was conducted in Medline and Web of Science databases using the Web of Knowledge interface, searching for the following terms:
 Topic=(health OR nutrition OR  medic$^*$)
AND Topic=(statistics analysis $^*$ OR regression model$^*$ OR
BMI data analysis$^*$ OR obesity$^*$).
The search was restricted to research articles in English. No date restrictions were applied.

We excluded observational studies investigating basic medical questions without explicit data analysis and statistical modelling.

The search identified more than 100 relevant papers. We classified these according to research question and modelling method. 
Rather than enumerate all papers, we identified the range of methods and selected papers accordingly to illustrate the methods, as described below. Where possible, we also mention software packages for {\it R} to ease the application of statistical models.

The paper is structured as follows. Sections 2, 3 and 4  review the models  and methods used to explore the link between BMI and other factors in terms of the summary statistics ({\it mean}, {\it probability} and {\it quantile}).  Section 5 discusses methods for dealing with data quality issues such as non-response, missing and outliers. Section 6 outlines BMI-based complex data analysis methods. Section 7 concludes.

\section{Mean methods \label{Sec2}}

In this section, we describe mean-based correlation measurements such as the Pearson's correlation coefficient, and mean-based regression models such as linear regression models.
The linear regression models here include a simple multiple regression, linear mixed models or linear time series models.

\subsection{Influences of BMI: correlation analysis}

The population correlation coefficient $corr(X, Y)$  between two random variables X and Y with expected values $E(X)$ and $E(Y)$ can be applied for the simple dependency measurement of BMI and other factors. This measurement, however, needs to take into account potential time-variation and repeated measurement of BMI data. Component analysis based dependency measurers can be used for this purpose. For example, \cite{tangugsorn2001obstructive} used  canonical-correlation analysis (CCA) to demonstrate the relationship of cervico-craniofacial skeletal and upper airway soft tissue morphology to comprehend the complicated pathogenic components in obese  $(BMI >= 30 kg/m^2)$  and non-obese $(BMI < 30 kg/m^2)$  patients. \cite{hu2000effects} used CAA to examine the relationship between obesity, body fat distribution and lipoprotein profiles. 

CAA is usually defined as follows. Given two sets of random variables such as $ X = (x_1, \dots, x_n)'$ and $Y = (y_1, \dots, y_m)'$, CAA is appropriate to measure the associations between $X_i$ and $Y_j$. It can be used to identify the linear combinations with  maximum correlation between the factors, extending bivariate correlation analysis 
while aiming at (i) dimension reduction by means of selecting only the best components and (ii) solving collinearity issues between highly correlated variables via the use of (independent) predictors. R-packages {\it CCA} 
and {\it candisc} 
can be used in numerical analysis.

Other modern dependency measurements such as the Functional Singular Component Analysis developed by \cite{yang2011functional} have been used in quantifying the dependency between BMI and systolic blood pressure (SBP) as an indicator of the general individual health status.

\subsection{Dietary effect: linear regression modelling}

Epidemiologic studies have demonstrated significant associations between dietary factors and weight gain. Statistical model can provide further support and measure the strength of association.

Based on longitudinal data from a population-based mammography screening program introduced between 1987 and 1990  
in central Sweden, \cite{newby2006longitudinal} used a multiple linear regression model \[Y=\bbeta'\bmx+\epsilon,\] to estimate the associations between a change in BMI and changes in food patterns. 


Using a baseline dietary intake and data collected by food-frequency questionnaire, \cite{hughes2008higher} applied a linear mixed effect model to assess the change in BMI over time in the general population and the longitudinal relation between flavonoid intake and BMI at baseline. 

{\it R} packages such as {\it lme4}
are available to implement linear mixed-effects models.

\subsection{Infancy effect: a linear mixed-effects model approach}

While a genetic contribution to obesity susceptibility has been identified (\cite{frayling2007common}), the correlation between longer duration of breastfeeding and the fat mass- and obesity-associated (FTO) gene has been subject to further analysis. Based on cohort data for children who were followed up from birth to 14 years of age, \cite{abarin2012impact} first set the cut off to be age 1.5 years for all individuals to model BMI denoted as $Y$ in terms of $\bmx_r$ representing time-independent covariates, including two FTO genotypes:  
  \begin{displaymath}
   Y = \left\{
     \begin{array}{lr}
       \beta_{0,0}+\beta_{0,1}age+ \bmx_r \bbeta_r'&: age<1.5\\
       \beta_{1,0}+\beta_{1,1}age+ \bmx_r \bbeta_r'&: age \geq 1.5,
     \end{array}
   \right.
\end{displaymath}
where each of the models is a fixed effect model. BMI is continuous across all ages, so that the model with enforced continuity at the cut off may best match the BMI process and then have an equation as a mixed effect model \[Y=\beta_0+\beta_1 (age-1.5)+I_{(age<1.5)}\beta_2^{infant}(age-1.5)^2+I_{(age \geq 1.5)}\beta_2^{child}(age-1.5)^2\]
\[+\bmx_r' \bbeta_r+\epsilon.\]
The intercept term $\beta_0$ here represents a random effect intercept due to the heterogeneity of age groups or the difference in BMI growth trajectories across individuals. Accordingly, both  $\beta_1$ and $\beta_2$ are the random effect coefficients. Let $\bgamma=(\beta_0, \beta_1, \beta_2^{infant}, \beta_2^{child})'$ be a collection of these random coefficients then these first four terms of the model can be written as $\bgamma' \bmz$. Hence, the model above is a standard linear mixed model which is a special case of model (1):
\begin{equation}
Y=\bgamma' \bmz+ \bbeta'\bmx+\epsilon,
\end{equation}
where  $\bbeta=\bbeta_r$ and $\bmx=\bmx_r$ stand for the fixed effects term. We assume that $E(\epsilon)=0$ and variance-covariance matrix $Var(\epsilon)=G$ is known.


Similarly, \cite{warrington2013modelling} used a linear mixed-effects approach to model BMI trajectories in children for genetic association studies. Comparing four different mixed-effect models for their data from the Western Australian Pregnancy Cohort they found that the semiparametric linear mixed model was the most efficient for modelling childhood growth to detect modest genetic effects in this cohort. Again, the semiparametric and nonparametric mixed-effects models are special cases of model (1). They have the general forms as
\[Y=X\,\beta+Z\, u+ f_3(t)+\epsilon, \]
and
\[Y=f_1(X\,\beta)+f_2(Z\, u)+f_3(t)+\epsilon, \]
respectively where $f_1(.)$ and $f_2(.)$ are nonparametric and non-random functions and $f_3(.)$ specifies a random function.


\section{Probability models\label{Sec3} }

\subsection{Logistic regression}

Logistic regression is a type of analysis used for predicting the outcome of a categorical dependent variable based on one or more predictor variables. For example, logistic regression can be used to explore the  association of BMI with diabetes risk, setting the categorical dependent variable $Y$ as binary, $Y=1$ for diabetes, and $Y=0$ otherwise. For example, \cite{boffetta2011body} examined the association in logistic regression models
\[logit(P[Y=1])=\log(\frac{P[Y=1]}{P[Y=0]})=\beta_0+\beta_1\,x_1+...+\beta_m\,x_m,\]
by employing BMI as an independent categorical variable: ten BMI {\it categories} were established. The categories were chosen to improve the ability to investigate the association between BMI and diabetes, in particular at the extremes of the BMI distribution.

\cite{razak2013change} examined the change in BMI across all segments of the BMI distribution in 96 countries, and assessed whether the BMI distribution is changing between cross-sectional surveys conducted at different time points. As the number of survey cycles per country varied between two 
and five 
, they use multilevel regression models, between countries and within countries over survey cycles.
Multilevel models are particularly appropriate here as the research design and data collection are organized at more than one level. 

A logistic multilevel regression such as \[\log(\frac{p_{ij}}{1-p_{ij}})=\beta\, x_{ij}+ u_j Z_{ij},\]
can be used to analyze data for participants that are organized at more than one level, where one considers a level-1 outcome, $Y_{ij}$, taking on a value of 1 with conditional probability $p_{ij}$, and $u_j$ is a random effect across level 2 units.


\subsection{Probabilistic index models for BMI analysis}

The typical properties of BMI distributions such as non-normal shape, skewness and long tail motivate many modern statistical methods for data analysis. For example, individual BMI may be affected by several risk factors and the mean, skewness and shape of BMI distributions may change with covariate patterns (\cite{beyerlein2008alternative}, \cite{beyerlein2008breastfeeding}).
Probabilistic index models (PIMs) have been proposed \citep{thas2012probabilistic} as a semiparametric framework for modelling the probabilistic index (PI) as a function of covariates.  PIMs summarize the covariate effects on the shape of the response distribution, while providing sufficient information on the covariate effect sizes. For the model, $Y$ and $Y^*$ are independent random response variables associated with covariate patterns $X$ and $X^*$, respectively, where $(X, Y )$ and $(X^*, Y^*)$ denote two independent and identically distributed random vectors. The PIM is denoted as $Pr[Y<Y^*|X, X^*]=g^{-1}(Z^T \bbeta)$ where $g$ is a link function defining the relationship between the PI and a linear predictor. $Z$ is a vector that contains elements from $X$ and $X^*$, $Z = X-X^*$, where $X$ and $X^*$ are 0/1 dummies coding for two distinct groups of the population. The PI is subject to the Wilcoxon-Mann-Whitney test assessing if the population distributions are identical without assuming normal distribution. Given the flexibility of the model, covariate vector could be specified in various ways. In particular, the response variables could be defined on an ordered scale, which can be discrete or continuous, for which the mean of the difference $Y -Y^*$ did not have a proper interpretation as an effect size, but for which the PI did. Thus, the PIM may be the preferred method whenever the PI is considered as a meaningful parameter for quantifying effect sizes.

\section{Quantile regression methods \label{Sec4}}

Mean regression based approaches such as multiple regression may explore the average effect effectively, for example in a change in mean BMI, 
but may not be able to identify how extreme values depend on or affect other issues of interest. Therefore, mean regression based methods may not be able to answer how food patterns may affect large changes in BMI or extreme BMI values or how obesity increases the likelihood of various diseases because the distribution of BMI data is typically skewed.

\cite{beyerlein2008alternative} employed different regression approaches to predict childhood BMI by using parental socio-demographic and lifestyle information as well as child data on sex and age. 
Comparison of generalized linear models (GLMs), quantile regression (QR) and generalized additive models for location, scale and shape (GAMLSS) by goodness-of-fit measures showed that quantile regression may be an effective approach to explore complex distributions (\cite{beyerlein2008alternative}).

A good introduction to QR can be found in \cite{koenker2001quantile}  and \cite{yu2003quantile}.
Assume a random variable $Y$ with a cumulative distribution function $F$. Given a probability level $\alpha \in [0,1]$, the $\alpha$th quantile of $Y$, denoted as $Q(\alpha)$, is defined as
\[
Q(\alpha)=\inf\{y:F(y)\geq \alpha \}.
\]
The $\alpha$th population quantile can also be obtained via the following optimization problem:
\[
\min_{Y} \sum_i\ \rho_\alpha  (Y_i - Q(\alpha))  ,
\]
where $\rho_\alpha (u)=u (\alpha- I[u \leq 0])$ ($I[\cdot ]$ indicates the indicator function) is the so-called check function.

Similar to standard ordinary least squares (OLS) approaches, QR models provide the possibility to examine the central tendency in BMI data as well as associations with covariates such as age. Additionally, estimation via QR allows to analyse how a particular percentile of the BMI distribution is associated with covariates. Hence, the advantage of QR is that, first, it can help to identify lower and upper extremes of BMI, which relate directly to underweight or overweight parts of the population. Secondly, it can be useful in deriving  BMI charts helping researchers and a lay audience to get a better understanding of the BMI distribution within the general population.

\subsection{Quantile regression: selected examples}


Given specific research questions, mean-based regression analyses may not be suitable to address specific problems of interest or may provide inconsistent estimates \citep{bottai2013use}. For example, studies on the relationship between sleep duration and BMI applying mean-based regression methods have shown inconsistent results. \cite{chen2012quantile}  re-examined the relationship by using quantile regression to account for the potential heterogeneous effect of sleep duration on BMI in different BMI categories and compared estimation results from different types of models. 

In a similar study, \cite{yang2013stochastic} used quantile regression to analyze 
the relationships between sleep, stress, and obesity by gender. They found that the relationships between BMI and covariates were not constant across the BMI distribution and between women and men. 

In the following part of this section we review statistical models of quantile regression used for BMI studies. We loosely divide the models into three categories: parametric, nonparametric and semiparametric.
All of these models can be regarded as specific cases of equation (1) with $m(Y)=Q(Y)$.

\subsection{Parametric QR for BMI}

The framework provided by the conditional quantile regression method has largely extended the scope of BMI analyses due to its flexibility of incorporating different covariates. There have been various studies investigating the relationship linking BMI with a number of external factors, ranging from biological measurements to social and economic effects. A parametric regression such as a multiple linear regression has some intuition for the interpretation of regression estimates, and can be fitted by the above optimization problem via {\it R} package {\it quantreg}.

Let $X=\{X_1, X_2,...\}$ be the vector of covariates. The $\alpha$th conditional quantile of $Y$ given $X$ is
\[
Q_Y(\alpha|X)=X'\beta(\alpha),
\]
where $\beta(\alpha)$ is the regression coefficient. $\beta(\alpha)$ can be estimated by solving
\begin{equation}
\hat{\beta}(\alpha)=\mathrm{argmin}_\beta \sum_i   \rho_\alpha  (Y_i - X_i'\beta) . \label{regressionquantile}
\end{equation}

\cite{costa2009decomposing} studied the cross-country gap in BMI between Italy and Spain in 2003 by applying a decomposition methodology to the entire BMI distribution using a quantile regression approach. They considered the steady-state BMI equilibrium
\[
\text{BMI}= x'\beta + f'\gamma + e'\delta + \epsilon,
\]
where $f$ is vector of food consumption measures, $e$ represents covariates of physical activity, $x$ is a vector of individual characteristics, and $\epsilon$ represents residuals due to unobserved effects.

The model is a sum of three regression terms $x'\beta$, $f'\gamma$ and $e'\delta$, which is also named {\it additive models}. That is, besides the standard regression effect of  individual characteristics, the method models other effects such as food consumption measures and physical activity via addition.
\cite{fenske2011identifying} extended the QR framework to an additive model to include $k$ additive effects, which allows the inclusion of nonlinear effects
\[
Q_Y(\alpha|X,Z)= X'\beta + \sum_{i=1}^k f_{\alpha,i}(Z),
\]
where $Z$ is a vector of additional variables of nonlinear terms, $f_{\alpha,i}$, for $i=1,...,k$, denotes generic functions of $Z$ with  nonlinear relationship. The $X$ here consider six effects: child's age, duration of breastfeeding, maternal BMI, maternal age, years of maternal education, years of education of the mother's partner. Other individual factors that have been included as additive terms in BMI studies include education, marital status, work stress, behaviours such as smoking and breakfast, diet, etc. Socio-economic factors such as income, health insurance, employments, etc. are also included in a number of analyses. For example, \cite{sturm2005body} examined the association of BMI among US elementary school children with food price and restaurant density. \cite{stifel2009childhood} used a quantile regression approach to explore the correlates of childhood overweight and ethnicity, gender and other influences in the United States. \cite{popkin2010recent} found a parametric QR with actual BMI data regressed against age and age squared to be the best fit to provide estimates for relations between age and BMI for upper extremes of the BMI distribution.

\subsection{Nonparametric QR for BMI}
If a parametric model is misspecified, estimates will be inconsistent and biased resulting in potentially misleading conclusions.  Non-parametric regression models, on the other hand, offer a more flexible way of modelling a relationship than parametric models, but may require more sophisticated methods and large sample sizes. 
For example, a nonparametric model for BMI would imply that the relationship between BMI and other covariates is unknown but assumed to follow an estimated smooth function. Fitting a smoothing quantile function could be done using a spline according to \cite{koenker1994quantile} and {\it R} function {\it rqss}. Alternatively, \cite{li2010application} considered a nonparametric model for age-specific BMI that uses a double-kernel-based method and an automatic bandwidth selection procedure. The method employs the basic settings of a double-kernel estimator from \cite{yu1998local}, which uses two local-linear kernels to smooth both variables $Y=BMI$ and $T=age$  with some adaptation for the survey data. Different from spline smoothing, kernel smoothing is a weighted average of all data points around a local area, where the weights are specified using a standard probability function as the kernel function and using a bandwidth to control the local area.

\subsection{Semiparametric QR for BMI}
\cite{heagerty1999semiparametric} considered a semiparametric model for age-specific BMI. Their model is based on the following linear representation of BMI:
\[
Y=\mu(T)+\sigma(T)\epsilon(T),
\]
where $T$ could be age or time or a continuous variable, $\mu(T)$ and $\sigma(T)$ are the location and scale functions, and $\epsilon(T)$ is the function of the noise term that depends on $T$.
 Under this model the $\alpha$th conditional quantile of $T$ is defined as
\[
Q(\alpha|T)=\mu(T)+\sigma(T)z(\alpha|T),
\]
where $z(\alpha|T)$ is the $\alpha$th quantile of $\epsilon(T)$.  This model is specified as a parametric model in terms of $\mu(T)$ and $\sigma(T)$, but nonparametric smoothing methods mentioned in Section 4.3 are used to fit both $\mu(T)$ and $\sigma(T)$. 
An alternative and popular semiparametric quantile regression model uses a normal transformation-based approach.

In order to deal with non-normal and non-homogeneous distributions among different age groups and to model the relationship between BMI and $T=age$, one of the most successful and most widely applied methods is the LMS (lambda-mu-sigma) model introduced by Cole and his collaborators \citep{cole1988fitting,cole1990lms,cole1992smoothing}. The LMS model uses an age-specific Box-Cox power transformation to yield normality. Let $\mu(T)$ and $\sigma(T)$ be the age-specific mean and standard deviation, and $\lambda(T)$ the Box-Cox power, then the age-dependent $\alpha$th regression quantile of the BMI distribution is given by
\begin{equation}
Q(\alpha|T)=\mu(T)\left(1+ \Phi^{-1}(\alpha) \lambda(T)\sigma(T)\right)^{1/\lambda(T)}, \label{LMS}
\end{equation}
where $Q(\alpha|T)$ is the conditional $\alpha$th quantile given $T$; $\Phi^{-1}(\cdot)$ is the inverse of standard normal distribution. Then smoothing functions $\mu(T)$, $\sigma(T)$  and $\lambda(T)$ are fitted nonparametrically.

Cole et al. have implemented the LMS method in several fundamental BMI studies 
(\cite{cole1995body}, \cite{cole2000establishing} and \cite{cole2007body}). The LMS method has been extensively used in weight-related research, and has become a `standard' framework for studying age-specific BMI or other growth references, along with an application in plotting charts for BMI or other growth references. For example, \cite{ogden2002centers} compare the U.S. Centers for Disease Control and Prevention (CDC) growth data from 2000 with historical data from 1977 using the LMS method. \cite{onis2007development} consider the LMS method for the development of BMI cut-offs for both children and adults.

\subsection{Density regression approach}
\cite{dunson2007bayesian} proposed a density regression to study the association between Luteinizing hormone (LH) and BMI in randomly selected women 
defined by:
\[f(y|\bmx)=\sum_{k=1}^{\infty} w_k(\bmx) N(\bmx' \bbeta, \sigma^2).\]
The aim of this study was to identify how changes in LH may affect the BMI distribution while adjusting for the potentially confounding effect of age. 
The conditional BMI distribution was not assumed to be normally distributed, but could be regarded as a mixture of conditional normal densities because mixtures of a sufficiently large number of normal densities can be used to approximate any smooth density accurately. Where the weights $w_k$ could or could not depend on $\bmx$, and could be inference by classical methods or Bayesian methods, although
\cite{dunson2007bayesian} assumed their dependency of $\bmx$ and used nonparametric Bayesian inference. 

\section{Data quality issues \label{Sec5}}
 Another challenge in the analysis of BMI data may be caused by data quality issues such as missing values, non-response and outliers in variable values. No statistical model can yield valid estimates while ignoring these issues. The following section describes four examples of appropriate statistical methods to deal with data quality problems.

\subsection{Multiple imputation methods}
\cite{elliott2007using} used survey data and a mixture model based multiple imputation to obtain the BMI distribution for a paediatric population in the presence of clerical errors.

The mixture model is defined by latent classes that have common
means, conditional on age and health centre to accommodate the disproportional sample
design, but differing covariances. The �clerical error class� is the class with the largest
covariance matrix determinant.

The general approach for multiple imputation is as follows. For the $i$th individual, $i=1 , . . ., n,$ let $\bmz_i$ be a $q-$dimensional outcome of interest. Each individual value is assumed to depend on a set of $p$ covariates $\bmx_i$. The associated covariance is given by the individual's latent variance class membership, which is denoted by the unobserved latent variable $C_i$, where $C_i=K$ indicates that the $i$th individual belongs to the clerical error class with the largest variability. The $C_i$ can be regarded as missing for all subjects belonging to the clerical error class. 
The complete-data  mixture model considered is
\begin{eqnarray}
\bmz_i|C_i&=&k \sim N_q(\mu_i, \, \Sigma_k), \nonumber \\
C_i &\sim& MULTI(1, p_i, ..., p_k) \nonumber
\end{eqnarray}
Once priors are postulated for model parameters $\bbeta$, $\Sigma_k$ and others such as $(p_1, ..., p_K) \sim DIRICHLET (1, 1, ..., 1)$, proper Bayesian inference such as Gibbs sampler can be applied.  All missing values are then replaced by imputed values from the Gibbs sampling procedure.

\subsection{Ignorable likelihood method}

In a study by \cite{little2011subsample}, 
some variables of interest, including BMI, were subject to missing data. 
They applied an ignorable likelihood (IL) method to multiply impute the full sample, and then used it for regression analyses. Results for different imputing methods were compared and yielded similar estimates for the effect of household income and education on blood pressure. 

In general, the IL approach requires a model for the distribution of covariates $W$ and an outcome variable $Y$, both with missing values, given a fully observed covariate set $Z$, indexed by parameters $\theta$, for example $p(w_i, y_i|z_i, \theta)$, where fully observed covariates can be treated as fixed (Little and Rubin (2002)). Integrating the missing variables out of the joint distribution and treating $\theta$ as the argument of the resulting density yields the IL
\[L_{ign}(\theta) =  \propto \prod_{i=1}^n p(w_{obs, i}, y_{obs, i}|z_i, \theta),\]
where $(w_{obs, i}, y_{obs, i})$ are the observed components of $(w_i, y_i)$ .

\subsection{Two-stage cluster samples for non-response}

Data may be missing due to non-response. \cite{rubin1976inference}  and \cite{little1994class} classified non-response mechanisms into three types: missingness completely at random (MCAR), when the probability of the non-response does not depend on clusters or survey variables; missingness at random (MAR), when
the probability of response depends only on the observed values; and non-ignorable non-response, when the probability of non-response depends on unobserved values. In their analysis, \cite{yuan2007model} dealt with a unit non-response rate about 40\% when households failed to answer questions in a questionnaire. 
To assess the relationship between cluster response rates and cluster means, they
plotted cluster sample response rates against cluster sample means of log(BMI), which displayed a
slightly linear trend with a correlation coefficient of 0.32, suggesting that the non-response mechanism
is not missingness completely at random (MCAR), and a cluster-specific non-ignorable (CSNI) non-response   mechanism may be indicated.

Based on a logarithm transformation of BMI measurements, they proposed several model-based estimates of the finite population mean for two-stage samples with unit non-response and compared them with existing methods by a simulation study. These models include: (1) applying standard two-stage mean estimators from complete response observations \citep{horvitz1952generalization} to non-response observations; (2) discarding non-respondents and basing estimates on predictions from a random-effects model fitted to respondents; (3) adding non-informative priors for the fixed parameters of a random-effects model and simulating draws from the posterior distribution of the parameters.

\subsection{ Bayesian method}

In order to deal with serious non-response and selection bias 
due to missing BMI values for a considerable number of individuals, \cite{nandram2010bayesian} used differential
probabilities for selection of these individuals 
A non-ignorable non-response model was proposed to estimate the finite population means of covariates where 
the log(BMI) values were used to obtain more normally distributed data. The model included a spline regression of log(BMI) on age, adjusted for several individual characteristics. 

Their data contained information on 
$N_i$ individuals for $i$ countries 
The authors assumed that the response indicator $r_{ij}$ for $j$th individual within the $i$th country related to BMI value $x_{ij}$ via
 \[r_{ij}|x_{ij}, \bbeta   \sim^{iid} Bernoulli \{exp(\beta_{0i}+\beta_{1i} x_{ij})/(1+exp(\beta_{0i}+\beta_{1i} x_{ij}))\},\]
 and employed a hierarchical setting for the regression parameters above.
 These $x_{ij}$ are a regression function of other covariates such as age, ethnicity and sex. Within the Bayesian framework, the sampled non-respondent BMI values are then obtained from their conditional posterior densities in the Metropolis-Hastings algorithm, and the non-sampled BMI values are drawn from their conditional posterior densities.

\section{ Data complexity issues \label{Sec6}}

Many scientific questions related to BMI can be answered by the analysis of cross-sectional data. 
However, BMI and associated 
factors may vary over time, resulting in repeated observations of the same variable at different time points. If longitudinal data are available, specific statistical methods or models are needed to take account of the nature of data and fully use the information provided.

Complex data issues may arise due to functional data such as longitudinal data, dimensional problems due to a high number of factors, and other big data issues. The following sections describe typical approaches used in this context. Statistical methods dealing with these types of complex data include dimensional reduction methods such as additive models, and variable selection methods such as lasso or group lasso.

\subsection{Generalized estimating equations (GEE)}

The generalized estimating equations (GEE), an extension of the quasi-likelihood approach, is a popular method to analyze longitudinal and other correlated data. Anthropometric measurements required for BMI-related research are naturally longitudinal observations correlated to other factors. For example, since the contribution of physical activity (PA) to the development of  BMI  in critical periods of childhood is of interest, \cite{remmers2013relationship} examined longitudinal relationships between PA and BMI z-scores by using GEE for the analysis.

GEE (\cite{liang1986longitudinal}) can be considered as a method for combining certain estimating equations in presence of time-dependent covariates.  Given a mean model, $m_{ij}$ subject to unknown parameter vector $\bbeta$, and a variance structure, $V_{i}$, the estimating equation is described as:
\[U(\bbeta) = \sum_{i=1}^N \frac{\partial \mu_{ij}}{\partial \bbeta_k} V_i^{-1} \{ Y_i - m_i(\bbeta)\}.\]
The parameter estimates solve $U(\bbeta)=0 $.

Applications of GEE in the analysis of BMI data take account of, for instance, the complexity of the data due to individual information being stratified by gender and baseline weight status (\cite{remmers2013relationship}, \cite{branum2011prepregnancy}).


\subsection{Generalized method of moments (GMM)}

\cite{lai2007marginal} analysed the relationship between BMI and future morbidity among children 
using longitudinal data with time-dependent covariates. They found that some of the estimating equations combined by GEEs with an independent correlation structure are not valid. The authors distinguished between three types of time-dependent covariates 
and provided a test for whether a time-dependent covariate is of a certain type. Results indicated that when a covariate is of type I or II, valid estimating equations are available that are not exploited by GEEs assuming an independent correlation structure. As a likelihood analysis is impossible or extremely difficult in this case, and to make optimal use of the valid estimating equations, they use the generalized method of moments (GMM) (\cite{hansen1982large}). The GMM requires just a partial specification of an estimating model and specifies a certain number of moment conditions. These moment conditions are functions of both the model parameters and the data, such that their expectation is zero at the true values of the parameters.


\subsection{Generalized additive model (GAM)}
Generalized additive models (GAM) combine properties of generalized linear models with additive models (\cite{hastie1990generalized}). Each additive term is typically modelled as nonparametric function.
\cite{gregory2011wages} used GAM to analyse how wages are affected by BMI and age. They model both BMI and wages nonparametrically by use of an oracle estimator. This study, however, focused on young workers and did not examine whether the effect of obesity changes as people age.  

\subsection{Variable selection }
Modern variable selection methods including ridge regression, bridge regression (\cite{frank1993statistical}), least absolute shrinkage and selection operator (LASSO) (\citep{tibshirani1996regression}), elastic net (\cite{zou2005regularization}) and clipped absolute deviation method (\cite{fan2001variable}) can be used for factor selection 
when covariates available for the analysis of BMI and the extent of their correlation may vary substantially between groups and settings.
The question then is which factors have the most significant effect on BMI and whether or not one factor dominates the contribution if the effect of those variables is estimated simultaneously. 

Studies using the lasso method have analysed, for example, if BMI has the most significant effect among several predictors for disease (progression) (\cite{hesterberg2008least}). 
Furthermore, in genome-wide association studies (GWAS) where measurements of individual BMI are observed repeatedly at various time points, the lasso technique can be applied for variable selection and shrinkage of longitudinal data sets. For this purpose, a sparse model is produced, in the sense that only those predictors for which regression coefficients are estimated as non-zero are retained in the final regression model, minimizing the usual sum of squared errors with a bound on the sum of the absolute values of the coefficients. By doing so, prediction accuracy can be improved, and the estimated model is more interpretable.

The {\it LASSO} technique is particularly suitable when the number of variables $p$ exceeds the number of observations $n$, i.e. where $p>n$ poses a problem for the regression analysis.
However, although Lasso is popular for its mathematical performance, it is not robust to skewed distributions. For highly skewed distributions, such as diabetes prevalence in the population, robust regression methods such as least absolute deviation (LAD) and the before mentioned quantile regression methods have received considerable attention recently in variable selection methods \citep{bradic2011penalized,wu2009variable}.

\section{Conclusions \label{Sec7}}

Obesity rates have been increasing over recent decades, causing significant concern among policy makers. Understanding which factors influence individual body weight and how exactly excess body fat is contributing to increased risk for disease may help to reduce the increased prevalence of several common disorders associated with obesity, thereby lessening the burden placed on health care systems.
Various statistical methods can be applied to guide effective public health action. Depending on the specific policy concern, research question or data available for analysis, both classical and modern methods can be used to improve the understanding of the complex system of intercorrelated influences on BMI. Since the choice of a specific method and its implementation may be challenging, this paper aimed to give an overview of available methods and guide the statistical analysis of BMI-related research.

\section{Acknowledgements}
We thank Dr Zhuo Sheng and Mr Nicola Attard-Montalto for collecting some references of this review. This work has been supported in part by the National Institute for Health Research Method Grant (NIHR-RMOFS-2013-03-09).

\bibliographystyle{abbrvnat}
\bibliography{bmi}
@article{heagerty1999semiparametric,
  title={Semiparametric estimation of regression quantiles with application to standardizing weight for height and age in US children},
  author={Heagerty, Patrick J and Pepe, Margaret S},
  journal={Journal of the Royal Statistical Society: Series C (Applied Statistics)},
  volume={48},
  number={4},
  pages={533--551},
  year={1999},
  publisher={Wiley Online Library}
}

@article{li2010application,
  title={Application of nonparametric quantile regression to body mass index percentile curves from survey data},
  author={Li, Yan and Graubard, Barry I and Korn, Edward L},
  journal={Statistics in medicine},
  volume={29},
  number={5},
  pages={558--572},
  year={2010},
  publisher={Wiley Online Library}
}

@article{cole1990lms,
  title={The LMS method for constructing normalized growth standards.},
  author={Cole, Tim J and others},
  journal={European journal of clinical nutrition},
  volume={44},
  number={1},
  pages={45--60},
  year={1990}
}

@inproceedings{chen2005growth,
  title={Growth Charts of Body Mass Index (BMI) With Quantile Regression.},
  author={Chen, Colin},
  booktitle={AMCS},
  pages={114--120},
  year={2005}
}

@article{stifel2009childhood,
  title={Childhood overweight in the United States: A quantile regression approach},
  author={Stifel, David C and Averett, Susan L},
  journal={Economics \& Human Biology},
  volume={7},
  number={3},
  pages={387--397},
  year={2009},
  publisher={Elsevier}
}

@article{beyerlein2008alternative,
  title={Alternative regression models to assess increase in childhood BMI},
  author={Beyerlein, Andreas and Fahrmeir, Ludwig and Mansmann, Ulrich and Toschke, Andr{\'e} M},
  journal={BMC medical research methodology},
  volume={8},
  number={1},
  pages={59},
  year={2008},
  publisher={BioMed Central Ltd}
}

@article{koenker1978regression,
  title={Regression quantiles},
  author={Koenker, Roger and Bassett, Gilbert},
  journal={Econometrica: journal of the Econometric Society},
  pages={33--50},
  year={1978},
  publisher={JSTOR}
}

@article{yu1998local,
  title={Local linear quantile regression},
  author={Yu, Keming and Jones, MC},
  journal={Journal of the American statistical Association},
  volume={93},
  number={441},
  pages={228--237},
  year={1998},
  publisher={Taylor \& Francis}
}

@article{wei2006quantile,
  title={Quantile regression methods for reference growth charts},
  author={Wei, Ying and Pere, Anneli and Koenker, Roger and He, Xuming},
  journal={Statistics in medicine},
  volume={25},
  number={8},
  pages={1369--1382},
  year={2006},
  publisher={Wiley Online Library}
}

@article{cole1992smoothing,
  title={Smoothing reference centile curves: the LMS method and penalized likelihood},
  author={Cole, Timothy J and Green, Pamela J},
  journal={Statistics in medicine},
  volume={11},
  number={10},
  pages={1305--1319},
  year={1992},
  publisher={Wiley Online Library}
}

@article{gannoun2002reference,
  title={Reference curves based on non-parametric quantile regression},
  author={Gannoun, Ali and Girard, St{\'e}phane and Guinot, Christiane and Saracco, J{\'e}r{\^o}me},
  journal={Statistics in medicine},
  volume={21},
  number={20},
  pages={3119--3135},
  year={2002},
  publisher={Wiley Online Library}
}

@article{herpertz2003secular,
  title={Secular trends in body mass index measurements in preschool children from the City of Aachen, Germany},
  author={Herpertz-Dahlmann, Beate and Geller, Frank and B{\"o}hle, Corinna and Khalil, Corinna and Trost-Brinkhues, Gabriele and Ziegler, Andreas and Hebebrand, Johannes},
  journal={European journal of pediatrics},
  volume={162},
  number={2},
  pages={104--109},
  year={2003},
  publisher={Springer}
}

@article{chen2012quantile,
  title={A quantile regression approach to re-investigate the relationship between sleep duration and body mass index in Taiwan},
  author={Chen, Chiang-Ming and Chang, Chen-Kang and Yeh, Chia-Yu},
  journal={International journal of public health},
  volume={57},
  number={3},
  pages={485--493},
  year={2012},
  publisher={Springer}
}

@article{costa2009decomposing,
  title={Decomposing body mass index gaps between Mediterranean countries: A counterfactual quantile regression analysis},
  author={Costa-Font, Joan and Fabbri, Daniele and Gil, Joan},
  journal={Economics \& Human Biology},
  volume={7},
  number={3},
  pages={351--365},
  year={2009},
  publisher={Elsevier}
}

@article{cole1994growth,
  title={Growth charts for both cross-sectional and longitudinal data},
  author={Cole, TJ},
  journal={Statistics in medicine},
  volume={13},
  number={23-24},
  pages={2477--2492},
  year={1994},
  publisher={Wiley Online Library}
}

@article{cole1988fitting,
  title={Fitting smoothed centile curves to reference data},
  author={Cole, TJ},
  journal={Journal of the Royal Statistical Society. Series A (Statistics in Society)},
  pages={385--418},
  year={1988},
  publisher={JSTOR}
}

@article{cole1995body,
  title={Body mass index reference curves for the UK, 1990.},
  author={Cole, Timothy J and Freeman, Jennifer V and Preece, Michael A},
  journal={Archives of disease in childhood},
  volume={73},
  number={1},
  pages={25--29},
  year={1995},
  publisher={BMJ Publishing Group Ltd and Royal College of Paediatrics and Child Health}
}

@article{cole2000establishing,
  title={Establishing a standard definition for child overweight and obesity worldwide: international survey},
  author={Cole, Tim J and Bellizzi, Mary C and Flegal, Katherine M and Dietz, William H},
  journal={Bmj},
  volume={320},
  number={7244},
  pages={1240},
  year={2000},
  publisher={BMJ}
}

@article{cole2007body,
  title={Body mass index cut offs to define thinness in children and adolescents: international survey},
  author={Cole, Tim J and Flegal, Katherine M and Nicholls, Dasha and Jackson, Alan A},
  journal={Bmj},
  volume={335},
  number={7612},
  pages={194},
  year={2007},
  publisher={BMJ}
}

@article{ogden2002centers,
  title={Centers for Disease Control and Prevention 2000 growth charts for the United States: improvements to the 1977 National Center for Health Statistics version},
  author={Ogden, Cynthia L and Kuczmarski, Robert J and Flegal, Katherine M and Mei, Zuguo and Guo, Shumei and Wei, Rong and Grummer-Strawn, Laurence M and Curtin, Lester R and Roche, Alex F and Johnson, Clifford L},
  journal={Pediatrics},
  volume={109},
  number={1},
  pages={45--60},
  year={2002},
  publisher={Am Acad Pediatrics}
}

@article{onis2007development,
  title={Development of a WHO growth reference for school-aged children and adolescents},
  author={Onis, Mercedes de and Onyango, Adelheid W and Borghi, Elaine and Siyam, Amani and Nishida, Chizuru and Siekmann, Jonathan},
  journal={Bulletin of the World Health Organization},
  volume={85},
  number={9},
  pages={660--667},
  year={2007},
  publisher={SciELO Public Health}
}

@article{sturm2005body,
  title={Body mass index in elementary school children, metropolitan area food prices and food outlet density},
  author={Sturm, Roland and Datar, Ashlesha},
  journal={Public health},
  volume={119},
  number={12},
  pages={1059--1068},
  year={2005},
  publisher={Elsevier}
}

@article{fenske2011identifying,
  title={Identifying risk factors for severe childhood malnutrition by boosting additive quantile regression},
  author={Fenske, Nora and Kneib, Thomas and Hothorn, Torsten},
  journal={Journal of the American Statistical Association},
  volume={106},
  number={494},
  year={2011}
}

@article{asirvatham2009examining,
  title={Examining diet quality and body mass index in rural areas using a quantile regression framework},
  author={Asirvatham, Jebaraj},
  journal={The Review of Regional Studies},
  volume={39},
  number={2},
  pages={149--169},
  year={2009}
}

@article{abarin2012impact,
  title={The impact of breastfeeding on FTO-related BMI growth trajectories: an application to the Raine pregnancy cohort study},
  author={Abarin, Taraneh and Wu, Yan Yan and Warrington, Nicole and Lye, Stephen and Pennell, Craig and Briollais, Laurent},
  journal={International journal of epidemiology},
  volume={41},
  number={6},
  pages={1650--1660},
  year={2012},
  publisher={IEA}
}

@article{branum2011prepregnancy,
  title={Prepregnancy body mass index and gestational weight gain in relation to child body mass index among siblings},
  author={Branum, Amy M and Parker, Jennifer D and Keim, Sarah A and Schempf, Ashley H},
  journal={American journal of epidemiology},
  volume={174},
  number={10},
  pages={1159--1165},
  year={2011},
  publisher={Oxford Univ Press}
}

@article{bayer2011factors,
  title={Factors Associated With Tracking of BMI: A Meta-Regression Analysis on BMI Tracking*},
  author={Bayer, Otmar and Kr{\"u}ger, Helia and Kries, R{\"u}diger and Toschke, Andr{\'e} M},
  journal={Obesity},
  volume={19},
  number={5},
  pages={1069--1076},
  year={2011},
  publisher={Wiley Online Library}
}

@article{dinsdale2011simple,
  title={A simple guide to classifying body mass index in children},
  author={Dinsdale, H and Ridler, C and Ells, LJ},
  journal={National Obesity Observatory, Oxford},
  year={2011}
}

@article{gregory2011wages,
  title={Wages, BMI, and Age: A Generalized Additive Model Using the Oracle Estimator},
  author={Gregory, Christian},
  journal={Available at SSRN 1975044},
  year={2011}
}

@book{hastie1990generalized,
  title={Generalized Adolitive Models},
  author={Hastie, Trevor J and Tibshirani, Robert John},
  number={43},
  year={1990},
  publisher={CRC Press}
}

@article{hermann2011association,
  title={The association of education with body mass index and waist circumference in the EPIC-PANACEA study},
  author={Hermann, Silke and Rohrmann, Sabine and Linseisen, Jakob and May, Anne M and Kunst, Anton and Besson, Herve and Romaguera, Dora and Travier, Noemie and Tormo, Maria-Jose and Molina, Esther and others},
  journal={BMC public health},
  volume={11},
  number={1},
  pages={169},
  year={2011},
  publisher={BioMed Central Ltd}
}

@article{hodge1993association,
  title={Association of body mass index and waist-hip circumference ratio with cardiovascular disease risk factors in Micronesian Nauruans.},
  author={Hodge, AM and Dowse, GK and Zimmet, PZ},
  journal={International journal of obesity and related metabolic disorders: journal of the International Association for the Study of Obesity},
  volume={17},
  number={7},
  pages={399},
  year={1993}
}

@article{dagan2013waist,
  title={Waist circumference vs body mass index in association with cardiorespiratory fitness in healthy men and women: a cross sectional analysis of 403 subjects},
  author={Dagan, Shiri Sherf and Segev, Shlomo and Novikov, Ilya and Dankner, Rachel and others},
  journal={Nutrition journal},
  volume={12},
  number={1},
  pages={12},
  year={2013},
  publisher={BioMed Central Ltd}
}

@article{eknoyan2008adolphe,
  title={Adolphe Quetelet (1796--1874)—the average man and indices of obesity},
  author={Eknoyan, Garabed},
  journal={Nephrology Dialysis Transplantation},
  volume={23},
  number={1},
  pages={47--51},
  year={2008},
  publisher={ERA-EDTA}
}

@article{elliott2007using,
  title={Using a mixture model for multiple imputation in the presence of outliers: the 'Healthy for life'project},
  author={Elliott, Michael R and Stettler, Nicolas},
  journal={Journal of the Royal Statistical Society: Series C (Applied Statistics)},
  volume={56},
  number={1},
  pages={63--78},
  year={2007},
  publisher={Wiley Online Library}
}

@article{hesterberg2008least,
  title={Least angle and l1 penalized regression: A review},
  author={Hesterberg, Tim and Choi, Nam Hee and Meier, Lukas and Fraley, Chris},
  journal={Statistics Surveys},
  volume={2},
  pages={61--93},
  year={2008},
  publisher={The author, under a Creative Commons Attribution License}
}

@article{janssen2004waist,
  title={Waist circumference and not body mass index explains obesity-related health risk},
  author={Janssen, Ian and Katzmarzyk, Peter T and Ross, Robert},
  journal={The American journal of clinical nutrition},
  volume={79},
  number={3},
  pages={379--384},
  year={2004},
  publisher={Am Soc Nutrition}
}

@article{larsson2008excess,
  title={Excess body fatness: an important cause of most cancers},
  author={Larsson, Susanna C and Wolk, Alicja},
  journal={The Lancet},
  volume={371},
  number={9612},
  pages={536--537},
  year={2008},
  publisher={Elsevier}
}

@article{little2011subsample,
  title={Subsample ignorable likelihood for regression analysis with missing data},
  author={Little, Roderick J and Zhang, Nanhua},
  journal={Journal of the Royal Statistical Society: Series C (Applied Statistics)},
  volume={60},
  number={4},
  pages={591--605},
  year={2011},
  publisher={Wiley Online Library}
}

@article{libayesian,
  title={Bayesian Group LASSO for Nonparametric Varying-Coefficient Models},
  author={Li, Jiahan and Li, Runze and Wu, Rongling},
  journal={Submitted}
}

@article{macinnis2006body,
  title={Body size and composition and prostate cancer risk: systematic review and meta-regression analysis},
  author={MacInnis, Robert J and English, Dallas R},
  journal={Cancer causes \& control},
  volume={17},
  number={8},
  pages={989--1003},
  year={2006},
  publisher={Springer}
}

@article{mandel2004height,
  title={Height-related changes in body mass index: a reappraisal},
  author={Mandel, Dror and Zimlichman, Eyal and Mimouni, Francis B and Grotto, Itamar and Kreiss, Yitshak},
  journal={Journal of the American College of Nutrition},
  volume={23},
  number={1},
  pages={51--54},
  year={2004},
  publisher={Am Coll Nutrition}
}

@article{nandram2010bayesian,
  title={A Bayesian analysis of body mass index data from small domains under nonignorable nonresponse and selection},
  author={Nandram, Balgobin and Choi, Jai Won},
  journal={Journal of the American Statistical Association},
  volume={105},
  number={489},
  year={2010}
}

@article{newby2006longitudinal,
  title={Longitudinal changes in food patterns predict changes in weight and body mass index and the effects are greatest in obese women},
  author={Newby, PK and Weismayer, Christoph and {\AA}kesson, Agneta and Tucker, Katherine L and Wolk, Alicja},
  journal={The Journal of nutrition},
  volume={136},
  number={10},
  pages={2580--2587},
  year={2006},
  publisher={Am Soc Nutrition}
}

@article{pfeffermann2007small,
  title={Small-area estimation under informative probability sampling of areas and within the selected areas},
  author={Pfeffermann, Danny and Sverchkov, Michail},
  journal={Journal of the American Statistical Association},
  volume={102},
  number={480},
  pages={1427--1439},
  year={2007},
  publisher={Taylor \& Francis}
}

@article{razak2013change,
  title={Change in the body mass index distribution for women: analysis of surveys from 37 low-and middle-income countries},
  author={Razak, Fahad and Corsi, Daniel J and Subramanian, SV},
  journal={PLoS medicine},
  volume={10},
  number={1},
  pages={e1001367},
  year={2013},
  publisher={Public Library of Science}
}

@article{renehan2008body,
  title={Body-mass index and incidence of cancer: a systematic review and meta-analysis of prospective observational studies},
  author={Renehan, Andrew G and Tyson, Margaret and Egger, Matthias and Heller, Richard F and Zwahlen, Marcel},
  journal={The Lancet},
  volume={371},
  number={9612},
  pages={569--578},
  year={2008},
  publisher={Elsevier}
}

@article{thas2012probabilistic,
  title={Probabilistic index models},
  author={Thas, Olivier and Neve, Jan De and Clement, Lieven and Ottoy, Jean-Pierre},
  journal={Journal of the Royal Statistical Society: Series B (Statistical Methodology)},
  volume={74},
  number={4},
  pages={623--671},
  year={2012},
  publisher={Wiley Online Library}
}

@article{tibshirani1996regression,
  title={Regression shrinkage and selection via the lasso},
  author={Tibshirani, Robert},
  journal={Journal of the Royal Statistical Society. Series B (Methodological)},
  pages={267--288},
  year={1996},
  publisher={JSTOR}
}

@article{thompson2003food,
  title={Food purchased away from home as a predictor of change in BMI z-score among girls},
  author={Thompson, Olivia M and Ballew, C and Resnicow, K and Must, A and Bandini, LG and Cyr, HDWH and Dietz, WH},
  journal={International journal of obesity},
  volume={28},
  number={2},
  pages={282--289},
  year={2003},
  publisher={Nature Publishing Group}
}

@article{torloni2009prepregnancy,
  title={Prepregnancy BMI and the risk of gestational diabetes: a systematic review of the literature with meta-analysis},
  author={Torloni, MR and Betr{\'a}n, AP and Horta, BL and Nakamura, MU and Atallah, AN and Moron, AF and Valente, O},
  journal={Obesity Reviews},
  volume={10},
  number={2},
  pages={194--203},
  year={2009},
  publisher={Wiley Online Library}
}

@article{bottai2013use,
  title={Use of quantile regression to investigate the longitudinal association between physical activity and body mass index},
  author={Bottai, Matteo and Frongillo, Edward A and Sui, Xuemei and O'Neill, Jennifer R and McKeown, Robert E and Burns, Trudy L and Liese, Angela D and Blair, Steven N and Pate, Russell R},
  journal={Obesity},
  year={2013},
  publisher={Wiley Online Library}
}

@article{tibshirani1996regression,
  title={Regression shrinkage and selection via the lasso},
  author={Tibshirani, Robert},
  journal={Journal of the Royal Statistical Society. Series B (Methodological)},
  pages={267--288},
  year={1996},
  publisher={JSTOR}
}

@article{bradic2011penalized,
  title={Penalized composite quasi-likelihood for ultrahigh dimensional variable selection},
  author={Bradic, Jelena and Fan, Jianqing and Wang, Weiwei},
  journal={Journal of the Royal Statistical Society: Series B (Statistical Methodology)},
  volume={73},
  number={3},
  pages={325--349},
  year={2011},
  publisher={Wiley Online Library}
}

@article{yang2013stochastic,
  title={Stochastic Variability in Stress, Sleep Duration, and Sleep Quality Across the Distribution of Body Mass Index: Insights from Quantile Regression},
  author={Yang, Tse-Chuan and Matthews, Stephen A and Chen, Vivian Y-J},
  journal={International journal of behavioral medicine},
  pages={1--10},
  year={2013},
  publisher={Springer}
}

@article{wu2009variable,
  title={Variable selection in quantile regression},
  author={Wu, Yichao and Liu, Yufeng},
  journal={Statistica Sinica},
  volume={19},
  number={2},
  pages={801},
  year={2009}
}

@article{wang2007obesity,
  title={The obesity epidemic in the United States—gender, age, socioeconomic, racial/ethnic, and geographic characteristics: a systematic review and meta-regression analysis},
  author={Wang, Youfa and Beydoun, May A},
  journal={Epidemiologic reviews},
  volume={29},
  number={1},
  pages={6--28},
  year={2007},
  publisher={Soc Epidemiolc Res}
}

@article{boffetta2011body,
  title={Body mass index and diabetes in Asia: a cross-sectional pooled analysis of 900,000 individuals in the Asia cohort consortium},
  author={Boffetta, Paolo and McLerran, Dale and Chen, Yu and Inoue, Manami and Sinha, Rashmi and He, Jiang and Gupta, Prakash Chandra and Tsugane, Shoichiro and Irie, Fujiko and Tamakoshi, Akiko and others},
  journal={PLoS One},
  volume={6},
  number={6},
  pages={e19930},
  year={2011},
  publisher={Public Library of Science}
}

@article{popkin2010recent,
  title={Recent dynamics suggest selected countries catching up to US obesity},
  author={Popkin, Barry M},
  journal={The American journal of clinical nutrition},
  volume={91},
  number={1},
  pages={284S--288S},
  year={2010},
  publisher={Am Soc Nutrition}
}

@article{duncan2012racial,
  title={Racial differences in the built environment-body mass index relationship? A geospatial analysis of adolescents in urban neighborhoods},
  author={Duncan, Dustin T and Castro, Marcia C and Gortmaker, Steven L and Aldstadt, Jared and Melly, Steven J and Bennett, Gary G and others},
  journal={International journal of health geographics},
  volume={11},
  number={1},
  pages={11},
  year={2012},
  publisher={Springer}
}

@article{kline2008wages,
  title={The wages of BMI: Bayesian analysis of a skewed treatment--response model with nonparametric endogeneity},
  author={Kline, Brendan and Tobias, Justin L},
  journal={Journal of Applied Econometrics},
  volume={23},
  number={6},
  pages={767--793},
  year={2008},
  publisher={Wiley Online Library}
}
@article{togo2001food,
  title={Food intake patterns and body mass index in observational studies},
  author={Togo, P and Osler, Merete and S{\o}rensen, Thorkild IA and Heitmann, BL},
  journal={International journal of obesity},
  volume={25},
  number={12},
  pages={1741--1751},
  year={2001},
  publisher={Nature Publishing Group}
}

@article{jokela2012body,
  title={Body mass index and depressive symptoms: instrumental-variables regression with genetic risk score},
  author={Jokela, Markus and Elovainio, Marko and Keltikangas-J{\"a}rvinen, Liisa and Batty, G David and Hintsanen, Mirka and Sepp{\"a}l{\"a}, Ilkka and K{\"a}h{\"o}nen, Mika and Viikari, Jorma S and Raitakari, Olli T and Lehtim{\"a}ki, Terho and others},
  journal={Genes, Brain and Behavior},
  volume={11},
  number={8},
  pages={942--948},
  year={2012},
  publisher={Wiley Online Library}
}

@article{bassett2008walking,
  title={Walking, cycling, and obesity rates in Europe, North America, and Australia},
  author={Bassett Jr, David R and Pucher, John and Buehler, Ralph and Thompson, Dixie L and Crouter, Scott E},
  journal={J Phys Act Health},
  volume={5},
  number={6},
  pages={795--814},
  year={2008}
}

@article{marwaha2006study,
  title={A study of growth parameters and prevalence of overweight and obesity in school children from Delhi},
  author={Marwaha, Raman K and Tandon, Nikhil and Singh, Yashpal and Aggarwal, Rashmi and Grewal, Khushi and Mani, Kalaivani},
  journal={Indian pediatrics},
  volume={43},
  number={11},
  pages={943},
  year={2006},
  publisher={INDIAN PEDIATRIC}
}

@article{cole1999centiles,
  title={Centiles of body mass index for Dutch children aged 0-20 years in 1980-a baseline to assess recent trends in obesity},
  author={Cole, Timothy J and Roede, Machteld J},
  journal={Annals of human biology},
  volume={26},
  number={4},
  pages={303--308},
  year={1999},
  publisher={Informa UK Ltd UK}
}

@article{conway2003effectiveness,
  title={Effectiveness of the US Department of Agriculture 5-step multiple-pass method in assessing food intake in obese and nonobese women},
  author={Conway, Joan M and Ingwersen, Linda A and Vinyard, Bryan T and Moshfegh, Alanna J},
  journal={The American journal of clinical nutrition},
  volume={77},
  number={5},
  pages={1171--1178},
  year={2003},
  publisher={Am Soc Nutrition}
}

@article{conway2004accuracy,
  title={Accuracy of dietary recall using the USDA five-step multiple-pass method in men: an observational validation study},
  author={Conway, Joan M and Ingwersen, Linda A and Moshfegh, Alanna J},
  journal={Journal of the American Dietetic Association},
  volume={104},
  number={4},
  pages={595--603},
  year={2004},
  publisher={Elsevier}
}

@article{johnson1996comparison,
  title={Comparison of multiple-pass 24-hour recall estimates of energy intake with total energy expenditure determined by the doubly labeled water method in young children},
  author={Johnson, Rachel K and Driscoll, Patricia and Goran, Michael I},
  journal={Journal of the American Dietetic Association},
  volume={96},
  number={11},
  pages={1140--1144},
  year={1996},
  publisher={Elsevier}
}

@article{kubik2005schoolwide,
  title={Schoolwide food practices are associated with body mass index in middle school students},
  author={Kubik, Martha Y and Lytle, Leslie A and Story, Mary},
  journal={Archives of Pediatrics \& Adolescent Medicine},
  volume={159},
  number={12},
  pages={1111},
  year={2005},
  publisher={Am Med Assoc}
}

@article{remmers2013relationship,
  title={Relationship Between Physical Activity and the Development of BMI in Children.},
  author={Remmers, Teun and Sleddens, Ester and Gubbels, Jessica and de Vries, Sanne and Mommers, Monique and Penders, John and Kremers, Stef and Thijs, Carel},
  journal={Medicine and science in sports and exercise},
  year={2013}
}

@article{sundquist1998influence,
  title={The influence of socioeconomic status, ethnicity and lifestyle on body mass index in a longitudinal study},
  author={Sundquist, Jan and Johansson, Sven-Erik},
  journal={International journal of epidemiology},
  volume={27},
  number={1},
  pages={57--63},
  year={1998},
  publisher={IEA}
}

@article{sundquist1999cardiovascular,
  title={Cardiovascular risk factors and the neighbourhood environment: a multilevel analysis.},
  author={Sundquist, Jan and Malmstr{\"o}m, Marianne and Johansson, Sven-Erik},
  journal={International Journal of Epidemiology},
  volume={28},
  number={5},
  pages={841--845},
  year={1999},
  publisher={IEA}
}

@article{beyerlein2008breastfeeding,
  title={Breastfeeding and Childhood Obesity: Shift of the Entire BMI Distribution or Only the Upper Parts\&quest},
  author={Beyerlein, Andreas and Toschke, Andr{\'e} M and von Kries, R{\"u}diger},
  journal={Obesity},
  volume={16},
  number={12},
  pages={2730--2733},
  year={2008},
  publisher={Nature Publishing Group}
}

@article{ogden2002prevalence,
  title={Prevalence and trends in overweight among US children and adolescents, 1999-2000},
  author={Ogden, Cynthia L and Flegal, Katherine M and Carroll, Margaret D and Johnson, Clifford L},
  journal={JAMA: the journal of the American Medical Association},
  volume={288},
  number={14},
  pages={1728--1732},
  year={2002},
  publisher={Am Med Assoc}
}

@inproceedings{shen2006biometric,
  title={Biometric statistical study of one-lead ECG features and body mass index (BMI)},
  author={Shen, Tsu-Wang and Tompkins, Willis J},
  booktitle={Engineering in Medicine and Biology Society, 2005. IEEE-EMBS 2005. 27th Annual International Conference of the},
  pages={1162--1165},
  year={2006},
  organization={IEEE}
}

@article{cha2008replication,
  title={Replication of genetic effects of FTO polymorphisms on BMI in a Korean population},
  author={Cha, Seong W and Choi, Sun M and Kim, Kil S and Park, Byung L and Kim, Jae R and Kim, Jong Y and Shin, Hyoung D},
  journal={Obesity},
  volume={16},
  number={9},
  pages={2187--2189},
  year={2008},
  publisher={Nature Publishing Group}
}

@article{heo2003hierarchical,
  title={Hierarchical linear models for the development of growth curves: an example with body mass index in overweight/obese adults},
  author={Heo, Moonseong and Faith, Myles S and Mott, John W and Gorman, Bernard S and Redden, David T and Allison, David B},
  journal={Statistics in medicine},
  volume={22},
  number={11},
  pages={1911--1942},
  year={2003},
  publisher={Wiley Online Library}
}

@article{majer2012time,
  title={Time trends and forecasts of body mass index from repeated cross-sectional data: a different approach},
  author={Majer, Istvan M and Mackenbach, Johan P and Baal, Pieter HM},
  journal={Statistics in Medicine},
  year={2012},
  publisher={Wiley Online Library}
}

@article{mishra2004multiple,
  title={Multiple imputation for body mass index: lessons from the Australian Longitudinal Study on Women's Health},
  author={Mishra, Gita D and Dobson, Annette J},
  journal={Statistics in medicine},
  volume={23},
  number={19},
  pages={3077--3087},
  year={2004},
  publisher={Wiley Online Library}
}

@article{kim2013novel,
  title={A Novel Method for Classifying Body Mass Index on the Basis of Speech Signals for Future Clinical Applications: A Pilot Study},
  author={Kim, Jong Yeol},
  journal={Evidence-Based Complementary and Alternative Medicine},
  volume={2013},
  year={2013},
  publisher={Hindawi Publishing Corporation}
}

@article{mclaren2009social,
  title={Social class and BMI among Canadian adults: A focus on occupational prestige},
  author={McLaren, Lindsay and Godley, Jenny},
  journal={Obesity},
  volume={17},
  number={2},
  pages={290--299},
  year={2009},
  publisher={Wiley Online Library}
}

@article{graziano2011cardiovascular,
  title={Cardiovascular regulation profile predicts developmental trajectory of BMI and pediatric obesity},
  author={Graziano, Paulo A and Calkins, Susan D and Keane, Susan P and O'Brien, Marion},
  journal={Obesity},
  volume={19},
  number={9},
  pages={1818--1825},
  year={2011},
  publisher={Nature Publishing Group}
}

@article{freedman2013skinfolds,
  title={Skinfolds and coronary heart disease risk factors are more strongly associated with BMI than with the body adiposity index},
  author={Freedman, David S and Ogden, Cynthia L and Goodman, Alyson B and Blanck, Heidi M},
  journal={Obesity},
  volume={21},
  number={1},
  pages={E64--E70},
  year={2013},
  publisher={Wiley Online Library}
}

@article{silva2012high,
  title={Is high body fat estimated by body mass index and waist circumference a predictor of hypertension in adults? A population-based study},
  author={Silva, Diego Augusto Santos and Petroski, Edio Luiz and Peres, Marco Aurelio and others},
  journal={Nutrition journal},
  volume={11},
  number={1},
  pages={112},
  year={2012},
  publisher={BioMed Central Ltd}
}

@article{lai2007marginal,
  title={Marginal regression analysis of longitudinal data with time-dependent covariates: a generalized method-of-moments approach},
  author={Lai, Tze Leung and Small, Dylan},
  journal={Journal of the Royal Statistical Society: Series B (Statistical Methodology)},
  volume={69},
  number={1},
  pages={79--99},
  year={2007},
  publisher={Wiley Online Library}
}

@article{walsh2011loess,
  title={The loess regression relationship between age and BMI for both Sydney World Masters Games athletes and the Australian national population},
  author={Walsh, Joe and Climstein, Mike and Heazlewood, Ian Timothy and Burke, Stephen and Kettunen, Jyrki and Adams, Kent and DeBeliso, Mark},
  journal={International journal of biological and medical sciences},
  volume={1},
  number={1},
  pages={33},
  year={2011}
}

@article{who2004appropriate,
  title={Appropriate body-mass index for Asian populations and its implications for policy and intervention strategies.},
  author={WHO Expert Consultation},
  journal={Lancet},
  volume={363},
  number={9403},
  pages={157},
  year={2004}
}

@article{yuan2007model,
  title={Model-based estimates of the finite population mean for two-stage cluster samples with unit non-response},
  author={Yuan, Ying and Little, Roderick JA},
  journal={Journal of the Royal Statistical Society: Series C (Applied Statistics)},
  volume={56},
  number={1},
  pages={79--97},
  year={2007},
  publisher={Wiley Online Library}
}

@article{yang2011functional,
  title={Functional singular component analysis},
  author={Yang, Wenjing and M{\"u}ller, Hans-Georg and Stadtm{\"u}ller, Ulrich},
  journal={Journal of the Royal Statistical Society: Series B (Statistical Methodology)},
  volume={73},
  number={3},
  pages={303--324},
  year={2011},
  publisher={Wiley Online Library}
}

@article{yu2013pre,
  title={Pre-Pregnancy Body Mass Index in Relation to Infant Birth Weight and Offspring Overweight/Obesity: A Systematic Review and Meta-Analysis},
  author={Yu, Zhangbin and Han, Shuping and Zhu, Jingai and Sun, Xiaofan and Ji, Chenbo and Guo, Xirong},
  journal={PloS one},
  volume={8},
  number={4},
  pages={e61627},
  year={2013},
  publisher={Public Library of Science}
}
@article{mills2005predicting,
  title={Predicting body fat using data on the BMI},
  author={Mills, Terence C},
  journal={J Stat Edu},
  volume={13},
  number={2},
  pages={1--3},
  year={2005}
}

@article{frank1993statistical,
  title={A statistical view of some chemometrics regression tools},
  author={Frank, LLdiko E and Friedman, Jerome H},
  journal={Technometrics},
  volume={35},
  number={2},
  pages={109--135},
  year={1993},
  publisher={Taylor \& Francis Group}
}

@article{zou2005regularization,
  title={Regularization and variable selection via the elastic net},
  author={Zou, Hui and Hastie, Trevor},
  journal={Journal of the Royal Statistical Society: Series B (Statistical Methodology)},
  volume={67},
  number={2},
  pages={301--320},
  year={2005},
  publisher={Wiley Online Library}
}

@article{hansen1982large,
  title={Large sample properties of generalized method of moments estimators},
  author={Hansen, Lars Peter},
  journal={Econometrica: Journal of the Econometric Society},
  pages={1029--1054},
  year={1982},
  publisher={JSTOR}
}

@article{fan2001variable,
  title={Variable selection via nonconcave penalized likelihood and its oracle properties},
  author={Fan, Jianqing and Li, Runze},
  journal={Journal of the American Statistical Association},
  volume={96},
  number={456},
  pages={1348--1360},
  year={2001},
  publisher={Taylor \& Francis}
}

@article{beyerlein2008breastfeeding,
  title={Breastfeeding and Childhood Obesity: Shift of the Entire BMI Distribution or Only the Upper Parts\&quest},
  author={Beyerlein, Andreas and Toschke, Andr{\'e} M and von Kries, R{\"u}diger},
  journal={Obesity},
  volume={16},
  number={12},
  pages={2730--2733},
  year={2008},
  publisher={Nature Publishing Group}
}

@article{toschke2008risk,
  title={Risk factors for childhood obesity: shift of the entire BMI distribution vs. shift of the upper tail only in a cross sectional study},
  author={Toschke, Andr{\'e} M and von Kries, R{\"u}diger and Beyerlein, Andreas and R{\"u}ckinger, Simon},
  journal={BMC Public Health},
  volume={8},
  number={1},
  pages={115},
  year={2008},
  publisher={BioMed Central Ltd}
}

@article{little1994class,
  title={A class of pattern-mixture models for normal incomplete data},
  author={Little, Roderick JA},
  journal={Biometrika},
  volume={81},
  number={3},
  pages={471--483},
  year={1994},
  publisher={Biometrika Trust}
}

@article{rubin1976inference,
  title={Inference and missing data},
  author={Rubin, Donald B},
  journal={Biometrika},
  volume={63},
  number={3},
  pages={581--592},
  year={1976},
  publisher={Biometrika Trust}
}

@article{horvitz1952generalization,
  title={A generalization of sampling without replacement from a finite universe},
  author={Horvitz, Daniel G and Thompson, Donovan J},
  journal={Journal of the American Statistical Association},
  volume={47},
  number={260},
  pages={663--685},
  year={1952},
  publisher={Taylor \& Francis Group}
}

@article{liang1986longitudinal,
  title={Longitudinal data analysis using generalized linear models},
  author={Liang, Kung-Yee and Zeger, Scott L},
  journal={Biometrika},
  volume={73},
  number={1},
  pages={13--22},
  year={1986},
  publisher={Biometrika Trust}
}

@article{bays2007relationship,
  title={The relationship of body mass index to diabetes mellitus, hypertension and dyslipidaemia: comparison of data from two national surveys},
  author={Bays, Harold E and Chapman, RH and Grandy, S},
  journal={International journal of clinical practice},
  volume={61},
  number={5},
  pages={737--747},
  year={2007},
  publisher={Wiley Online Library}
}

@article{efron2004least,
  title={Least angle regression},
  author={Efron, Bradley and Hastie, Trevor and Johnstone, Iain and Tibshirani, Robert},
  journal={The Annals of statistics},
  volume={32},
  number={2},
  pages={407--499},
  year={2004},
  publisher={Institute of Mathematical Statistics}
}

@article{taber2012state,
  title={State Disparities in Time Trends of Adolescent Body Mass Index Percentile and Weight-Related Behaviors in the United States},
  author={Taber, Daniel R and Stevens, June and Poole, Charles and Maciejewski, Matthew L and Evenson, Kelly R and Ward, Dianne S},
  journal={Journal of community health},
  volume={37},
  number={1},
  pages={242--252},
  year={2012},
  publisher={Springer}
}

@article{chen2012quantile,
  title={A quantile regression approach to re-investigate the relationship between sleep duration and body mass index in Taiwan},
  author={Chen, Chiang-Ming and Chang, Chen-Kang and Yeh, Chia-Yu},
  journal={International journal of public health},
  volume={57},
  number={3},
  pages={485--493},
  year={2012},
  publisher={Springer}
}

@article{vogli2014economic,
  title={Economic globalization, inequality and body mass index: a cross-national analysis of 127 countries},
  author={Vogli, Roberto De and Kouvonen, Anne and Elovainio, Marko and Marmot, Michael},
  journal={Critical Public Health},
  volume={24},
  number={1},
  pages={7--21},
  year={2014},
  publisher={Taylor \& Francis}
}

@article{frayling2007common,
  title={A common variant in the FTO gene is associated with body mass index and predisposes to childhood and adult obesity},
  author={Frayling, Timothy M and Timpson, Nicholas J and Weedon, Michael N and Zeggini, Eleftheria and Freathy, Rachel M and Lindgren, Cecilia M and Perry, John RB and Elliott, Katherine S and Lango, Hana and Rayner, Nigel W and others},
  journal={Science},
  volume={316},
  number={5826},
  pages={889--894},
  year={2007},
  publisher={American Association for the Advancement of Science}
}

@article{warrington2013modelling,
  title={Modelling BMI Trajectories in Children for Genetic Association Studies},
  author={Warrington, Nicole M and Wu, Yan Yan and Pennell, Craig E and Marsh, Julie A and Beilin, Lawrence J and Palmer, Lyle J and Lye, Stephen J and Briollais, Laurent},
  journal={PloS one},
  volume={8},
  number={1},
  pages={e53897},
  year={2013},
  publisher={Public Library of Science}
}

@article{hughes2008higher,
  title={Higher dietary flavone, flavonol, and catechin intakes are associated with less of an increase in BMI over time in women: a longitudinal analysis from the Netherlands Cohort Study},
  author={Hughes, Laura AE and Arts, Ilja CW and Ambergen, Ton and Brants, Henny AM and Dagnelie, Pieter C and Goldbohm, R Alexandra and van den Brandt, Piet A and Weijenberg, Matty P},
  journal={The American journal of clinical nutrition},
  volume={88},
  number={5},
  pages={1341--1352},
  year={2008},
  publisher={Am Soc Nutrition}
}

@article{newby2006longitudinal,
  title={Longitudinal changes in food patterns predict changes in weight and body mass index and the effects are greatest in obese women},
  author={Newby, PK and Weismayer, Christoph and {\AA}kesson, Agneta and Tucker, Katherine L and Wolk, Alicja},
  journal={The Journal of nutrition},
  volume={136},
  number={10},
  pages={2580--2587},
  year={2006},
  publisher={Am Soc Nutrition}
}

@article{tangugsorn2001obstructive,
  title={Obstructive sleep apnea: a canonical correlation of cephalometric and selected demographic variables in obese and nonobese patients},
  author={Tangugsorn, Vivat and Krogstad, Olaf and Espeland, Lisen and Lyberg, Torstein},
  journal={The Angle Orthodontist},
  volume={71},
  number={1},
  pages={23--35},
  year={2001}
}

@article{hu2000effects,
  title={Effects of obesity and body fat distribution on lipids and lipoproteins in nondiabetic American Indians: The Strong Heart Study},
  author={Hu, Dongsheng and Hannah, Judy and Gray, R Stuart and Jablonski, Kathleen A and Henderson, Jeffrey A and Robbins, David C and Lee, Elisa T and Welty, Thomas K and Howard, Barbara V},
  journal={Obesity research},
  volume={8},
  number={6},
  pages={411--421},
  year={2000},
  publisher={Wiley Online Library}
}

@article{newby2006longitudinal,
  title={Longitudinal changes in food patterns predict changes in weight and body mass index and the effects are greatest in obese women},
  author={Newby, PK and Weismayer, Christoph and {\AA}kesson, Agneta and Tucker, Katherine L and Wolk, Alicja},
  journal={The Journal of nutrition},
  volume={136},
  number={10},
  pages={2580--2587},
  year={2006},
  publisher={Am Soc Nutrition}
}

@article{koenker2001quantile,
  title={Quantile regression: An introduction},
  author={Koenker, Roger and Hallock, Kevin},
  journal={Journal of Economic Perspectives},
  volume={15},
  number={4},
  pages={43--56},
  year={2001}
}

@article{yu2003quantile,
  title={Quantile regression: applications and current research areas},
  author={Yu, Keming and Lu, Zudi and Stander, Julian},
  journal={Journal of the Royal Statistical Society: Series D (The Statistician)},
  volume={52},
  number={3},
  pages={331--350},
  year={2003},
  publisher={Wiley Online Library}
}

@article{tibshirani1996regression,
  title={Regression shrinkage and selection via the lasso},
  author={Tibshirani, Robert},
  journal={Journal of the Royal Statistical Society. Series B (Methodological)},
  pages={267--288},
  year={1996},
  publisher={JSTOR}
}

@article{dunson2007bayesian,
  title={Bayesian density regression},
  author={Dunson, David B and Pillai, Natesh and Park, Ju-Hyun},
  journal={Journal of the Royal Statistical Society: Series B (Statistical Methodology)},
  volume={69},
  number={2},
  pages={163--183},
  year={2007},
  publisher={Wiley Online Library}
}

@article{majer2013time,
  title={Time trends and forecasts of body mass index from repeated cross-sectional data: a different approach},
  author={Majer, Istvan M and Mackenbach, Johan P and Baal, Pieter HM},
  journal={Statistics in medicine},
  volume={32},
  number={9},
  pages={1561--1571},
  year={2013},
  publisher={Wiley Online Library}
}

@article{finkelstein2012obesity,
  title={Obesity and severe obesity forecasts through 2030},
  author={Finkelstein, Eric A and Khavjou, Olga A and Thompson, Hope and Trogdon, Justin G and Pan, Liping and Sherry, Bettylou and Dietz, William},
  journal={American journal of preventive medicine},
  volume={42},
  number={6},
  pages={563--570},
  year={2012},
  publisher={Elsevier}
}

@article{rigby2005generalized,
  title={Generalized additive models for location, scale and shape},
  author={Rigby, RA and Stasinopoulos, DM},
  journal={Journal of the Royal Statistical Society: Series C (Applied Statistics)},
  volume={54},
  number={3},
  pages={507--554},
  year={2005},
  publisher={Wiley Online Library}
}

@article{lee1992modeling,
  title={Modeling and forecasting US mortality},
  author={Lee, Ronald D and Carter, Lawrence R},
  journal={Journal of the American statistical association},
  volume={87},
  number={419},
  pages={659--671},
  year={1992},
  publisher={Taylor \& Francis Group}
}

@article{frayling2007common,
  title={A common variant in the FTO gene is associated with body mass index and predisposes to childhood and adult obesity},
  author={Frayling, Timothy M and Timpson, Nicholas J and Weedon, Michael N and Zeggini, Eleftheria and Freathy, Rachel M and Lindgren, Cecilia M and Perry, John RB and Elliott, Katherine S and Lango, Hana and Rayner, Nigel W and others},
  journal={Science},
  volume={316},
  number={5826},
  pages={889--894},
  year={2007},
  publisher={American Association for the Advancement of Science}
}

@article{asirvatham2009examining,
  title={Examining diet quality and body mass index in rural areas using a quantile regression framework},
  author={Asirvatham, Jebaraj},
  journal={The Review of Regional Studies},
  volume={39},
  number={2},
  pages={149--169},
  year={2009}
}

@article{hansen1982large,
  title={Large sample properties of generalized method of moments estimators},
  author={Hansen, Lars Peter},
  journal={Econometrica: Journal of the Econometric Society},
  pages={1029--1054},
  year={1982},
  publisher={JSTOR}
}

@article{tibshirani1996regression,
  title={Regression shrinkage and selection via the lasso},
  author={Tibshirani, Robert},
  journal={Journal of the Royal Statistical Society. Series B (Methodological)},
  pages={267--288},
  year={1996},
  publisher={JSTOR}
}

@book{world2000obesity,
  title={Obesity: preventing and managing the global epidemic},
  author={World Health Organization},
  number={894},
  year={2000},
  publisher={World Health Organization}
}

@article{koenker1994quantile,
  title={Quantile smoothing splines},
  author={Koenker, Roger and Ng, Pin and Portnoy, Stephen},
  journal={Biometrika},
  volume={81},
  number={4},
  pages={673--680},
  year={1994},
  publisher={Biometrika Trust}
}

@article{ells2009preventing,
  title={Preventing childhood obesity through lifestyle change interventions. A briefing paper for commissioners},
  author={Ells, LJ and Cavill, N},
  journal={National Obesity Observatory, Oxford},
  year={2009}
}

\end{document}